\documentclass[%
reprint,
superscriptaddress,
 amsmath,amssymb,
 longbibliography,
prx,
]{revtex4-1}
\usepackage{textcomp}
\usepackage{gensymb} 
\usepackage{mathtools}
\usepackage{dsfont,physics,xcolor,comment}
\usepackage{graphicx,subfigure}
\usepackage{dcolumn}
\usepackage{bm}
\usepackage{hyperref}
\usepackage[capitalize]{cleveref}
\usepackage[mathlines]{lineno}

\def \todo #1{\textcolor{red}{#1}}
\def \mc #1{\mathcal{#1}}

\newcommand{\dlangle}{\ensuremath{\langle\!\langle}}
\newcommand{\drangle}{\ensuremath{\rangle\!\rangle}}
\newcommand{\dket}[1]{\lvert #1 \drangle}
\newcommand{\dbra}[1]{\dlangle #1 \rvert}

\makeatletter
\def\dketbra#1{\def\tempa{#1}\futurelet\next\dketbra@i}
\def\dketbra@i{\ifx\next\bgroup\expandafter\dketbra@ii\else\expandafter\dketbra@end\fi}
\def\dketbra@ii#1{\ensuremath{\dket{\tempa}\!\dbra{#1}}}
\def\dketbra@end{\ensuremath{\dket{\tempa}\!\dbra{\tempa}}}
\makeatother

\makeatletter
\def\dbraket#1{\def\tempa{#1}\futurelet\next\dbraket@i}
\def\dbraket@i{\ifx\next\bgroup\expandafter\dbraket@ii\else\expandafter\dbraket@end\fi}
\def\dbraket@ii#1{\ensuremath{\dbra{\tempa}#1 \drangle}}
\def\dbraket@end{\ensuremath{\dbra{\tempa}\tempa\drangle}}
\makeatother

\begin{document}

\preprint{APS/123-QED}

\title{Algorithmic cooling for resolving state preparation and measurement errors in quantum computing}
\author{Raymond Laflamme}
\affiliation{Institute for Quantum Computing and Department of Physics and Astronomy, University of Waterloo, Waterloo, Ontario N2L 3G1, Canada}
\affiliation{Perimeter Institute for Theoretical Physics, Waterloo, Ontario N2L 2Y5, Canada}

\author{Junan Lin}\email{junan.lin@uwaterloo.ca}
\affiliation{Institute for Quantum Computing and Department of Physics and Astronomy, University of Waterloo, Waterloo, Ontario N2L 3G1, Canada}

\author{Tal Mor}
\affiliation{Computer Science Department, Technion - Israel Institute of Technology, Technion city, Haifa 3200003, Israel}
\date{\today}

\begin{abstract}
State preparation and measurement errors are commonly regarded as indistinguishable. 
The problem of distinguishing state preparation (SPAM) errors from measurement errors is important to the field of characterizing quantum processors.
In this work, we propose a method to separately characterize SPAM errors using a novel type of algorithmic cooling protocol called measurement-based algorithmic cooling (MBAC).
MBAC assumes the ability to perform (potentially imperfect) projective measurements on individual qubits, which is available on many modern quantum processors.
We demonstrate that MBAC can significantly reduce state preparation error under realistic assumptions, with a small overhead that can be upper bounded by measurable quantities.
Thus, MBAC can be a valuable tool not only for benchmarking near-term quantum processors, but also for improving the performance of quantum processors in an algorithmic manner.

\end{abstract}

\pacs{Valid PACS appear here}
\maketitle

\section{Introduction}\label{sec_intro}
Quantum computers are believed to solve some problems more efficiently \cite{Deutsch1992,Grover1996}, and sometimes exponentially more efficiently \cite{Simon1997,Shor1999} than classical computers. 
On the other hand, various types of noise make it hard to design and build useful quantum computers and perform quantum algorithms. 
The main types of noise include state preparation errors, gates errors, dephasing and relaxation during computation, and measurements errors, along also with qubit-correlated errors.
Among these, state preparation and measurement (SPAM) errors are usually studied as a whole, because they are fundamentally hard to distinguish.
For example, if one tries to prepare a target state $\ket{0}$, immediately measures it and gets an outcome $\ket{1}$, it is not clear whether the state preparation or the measurement contributes more to the inconsistency.

Algorithmic cooling (AC), or more accurately heat-bath AC \cite{boykin2002algorithmic}, is a method that employs relaxation with a bath for cooling spins. 
We show here that an extended form of algorithmic cooling (AC), which we call measurement-based algorithmic cooling, can be a useful tool for distinguishing state preparation errors from measurement errors. 
Most commonly, various AC algorithms~\cite{Fernandez2004Algorithmic,Schulman2005Physical,Baugh2005Experimental,Ryan2008,Elias2011a,Brassard2014a,Park2015,Elias2011,Rodriguez-Briones2017,Rodriguez-Briones2017a} work under the assumptions that the system consists of qubits, and that the state preparation noise is thermal.
In addition, it is commonly assumed that the system is an ensemble of many identical subsystems, such as in NMR quantum computing. 
In ensemble quantum computing, some AC algorithms are sometimes performed prior to the standard quantum computing process, as part of a prolonged state preparation process.
These algorithms re-distribute the entropy among the qubits, so that the state of a subset of target qubits after AC is more pure, as if they were put in a lower temperature than the initial temperature of the original thermal state.

In the ensemble, a vital assumption is the limitation on the quantum computing abilities: it is assumed that operations cannot be applied onto individual qubits, but only to the same qubits in each of the many parallel quantum computers.
Consequently, one is limited to ensemble measurements that estimate expectation values of certain operators.
On the other hand, AC in other quantum computing systems can be less restrictive: if projective measurements are allowed, a trivial way to cool a qubit is to measure it.
If the outcome is $0$, the system is in the $\ket{0}$ state immediately after the measurement, corresponding to a zero temperature (discussed later).
If the outcome is $1$, then one can perform a Pauli X gate to flip it to $\ket{0}$, achieving the same result.
However, there are two major issues with this approach: first, in some existing gate-based quantum computing architectures, the measurement process only takes place at the final stage of the computation.
No computation can proceed on the measured qubit and a new round of computation must be initiated.
In other words, the measurement ``destroys'' the system of interest, so this naive cooling method is not so useful for doing further computations.
Second, this method works only if measurement errors are absent, and there is a perfect correspondence between the measurement result and the post-measurement state; if this is not true, we cannot safely deduce the post-measurement state.

In this work, we propose a novel AC protocol under the assumption that (potentially imperfect) measurements can be applied on individual qubits, which resolves the above issues.
We call this new class ``measurement-based AC'' (MBAC) since measurement is used as a resource to cool.
The outline of this paper is as follows.
In \cref{sec_review_AC} we give a review of conventional AC protocols, which motivate the new MBAC protocols.
In \cref{sec_mbac} we describe $k$-qubit MBAC assuming ideal measurements on the ancillary qubits, and compare it with the optimal reversible AC scheme.
In \cref{sec_SPAM_char} we describe a procedure to separately characterize SPAM errors using MBAC.
In \cref{sec_measurement_error} we relax the assumption made in \cref{sec_mbac} and analyze MBAC when the ancillary qubits have finite measurement errors, and derive a lower bound for its performance.
In \cref{sec_trials_needed} we study the number of trials needed to cool down a target qubit by a desired multiplicative factor, and illustrate the practical usefulness of MBAC despite its probabilistic nature.
Finally in \cref{sec_conclusion} we conclude our paper and point to some future research directions.

\section{Results}
\subsection{Review of conventional AC}\label{sec_review_AC}
We first describe AC protocols, notations and goals, in a way that leads more naturally to our novel type of AC, the MBAC.
Below, we will denote the Pauli matrices $X,Y,Z$ by $\sigma_{x}, \sigma_{y}, \sigma_{z}$, and the $2\times 2$ identity matrix by $\sigma_{I}$.
The ideal state preparation step should initialize a single qubit to the $\ket{0}$ state.
We assume that due to imperfect state preparation processes, a bit-flip (equivalently, $\sigma_{x}$) error occurs with probability $\delta$, so that the actual initial state is described by the following mixed state density matrix:
\begin{equation}\label{eqn_initial_state_defn}
    \rho = (1-\delta)\ketbra{0}{0} + \delta \ketbra{1}{1}.
\end{equation}
This represents a state preparation error on the quantum processor.
Throughout this work we will assume that $0\leq \delta < 1/2$, where $\delta=0$ corresponds to the pure state $\ket{0}$ and $\delta \rightarrow 1/2$ corresponds to the completely mixed state.

Algorithmic cooling procedures are designed to reduce $\delta$ on the target qubit towards 0.
The term ``cooling'' comes from viewing $\rho$ as a thermal state~\footnote{This correspondence can be made for all qubit density matrices that are diagonal in the eigenspace of the Hamiltonian, but not necessarily for higher dimensional systems.},
\begin{equation}\label{eqn_thermal}
    \rho = \frac{1}{Z} e^{-\beta \hat{H}} = \frac{1}{Z} \begin{pmatrix}
    e^{\frac{\hbar \omega}{2k_B T}} & 0\\
    0 & e^{-\frac{\hbar \omega}{2k_B T}} \end{pmatrix}
\end{equation}
where $\hat{H} = - \frac{1}{2}\hbar \omega \sigma_{z}$ is the bare qubit Hamiltonian, $k_{B}$ is the Boltzmann constant, $T$ is an effective temperature, and $Z$ is the partition function so that $\rho$ is normalized.
From \cref{eqn_initial_state_defn} and \cref{eqn_thermal} we can identify $\delta = (e^{\frac{\hbar \omega}{k_B T}}+1)^{-1}$.
Solving for $T$ gives $k_{B} T = \hbar \omega / \log(\frac{1}{\delta}-1)$, so that $T$ is closer to $0$ when $\delta$ is closer to $0$, for a fixed $\omega$.
Therefore, reducing its effective temperature increases its probability of being in the ground state.

AC has two main variants, namely, the Schulman-Vazirani reversible scheme~\cite{Schulman1999} and the heat-bath scheme~\cite{Fernandez2004Algorithmic,boykin2002algorithmic}. 
In the reversible variant, the relaxation time is assumed to be infinite or extremely large relative to the computing time. 
In the heat-bath variant, one assumes that in addition to computational (or, target) qubits that relax very slowly, there are also ancillary ``reset'' qubits that can relax back to thermal equilibrium much faster (through interaction with a heat-bath) than the target qubits.
We first review the heat-bath scheme using a minimal example with $m=3$ qubits~\cite{Fernandez2004Algorithmic}, which inspires our measurement-based protocol.
We will review the reversible variant later in \cref{sec_mbac} when we compare MBAC with conventional AC.

\begin{figure}[ht]
\centering
\includegraphics[width=0.35\columnwidth]{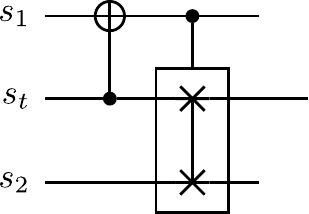}
\caption{The circuit for 3-qubit BCS aiming to cool down the target spin $s_t$. Within the box is a SWAP gate, which is controlled by spin $s_{1}$.}
\label{fig_BCS}
\end{figure}

\begin{figure}[ht]
\centering
\includegraphics[width=0.89\columnwidth]{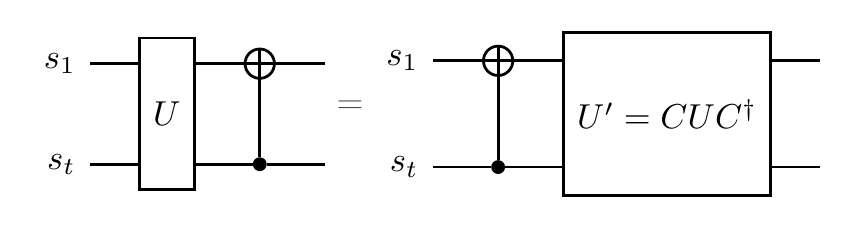}
\caption{Two equivalent circuits where the order of two unitary gates are exchanged, and the second gate on the RHS is replaced by the original gate conjugated by the first gate. $C$ stands for the CNOT gate in this case.}
\label{fig_BCS_detail}
\end{figure}

Let us now assume that there is a target spin to be cooled called $s_t$, and two ancillary spins $s_1$ and $s_2$.
Starting with the system in a state $\rho_1 \otimes \rho_t \otimes \rho_2$ where all $\rho$'s are described by \cref{eqn_initial_state_defn} with the same value of $\delta$, we can apply a basic compression subroutine (BCS) to reduce $\delta$ on $s_t$ (see \cref{fig_BCS}).
The BCS involves essentially two steps: first, a CNOT operation controlled by $s_t$ and targeted on $s_1$; second, a controlled-SWAP (CSWAP) operation controlled by $s_1$, and targeted on $s_t$ and $s_2$.
To see why this protocol works, first consider the evolution $s_{1}$ and $s_{t}$ after the CNOT gate.
We replace the noisy states of $s_{t}$ and $s_{1}$ by assuming an error model is applied onto pure qubit states.
The error model is such that a $\sigma_{x}$ channel occurs independently on each qubit after the preparation.
Referring to \cref{fig_BCS_detail}, this corresponds to the following four possibilities where $U$ equals to
\begin{enumerate}
    \item $\sigma_{I} \otimes \sigma_{I},\ p=(1-\delta)^{2}$;
    \item $\sigma_{I} \otimes \sigma_{x},\ p=\delta(1-\delta)$;
    \item $\sigma_{x} \otimes \sigma_{I},\ p=\delta(1-\delta)$;
    \item $\sigma_{x} \otimes \sigma_{x},\ p=\delta^{2}$;
\end{enumerate}
with the corresponding probabilities listed after each case.
In the above the left error-operator is applied onto $s_{1}$ and the right operator onto $s_{t}$.
From \cref{fig_BCS_detail} and the mapping rule of Pauli operators under the CNOT gate, the four cases can be written equivalently where $U'$ equals to
\begin{enumerate}
    \item $\sigma_{I} \otimes \sigma_{I},\ p=(1-\delta)^{2}$;
    \item $\sigma_{x} \otimes \sigma_{x},\ p=\delta(1-\delta)$;
    \item $\sigma_{x} \otimes \sigma_{I},\ p=\delta(1-\delta)$;
    \item $\sigma_{I} \otimes \sigma_{x},\ p=\delta^{2}$.
\end{enumerate}
Since $s_{1}$ and $s_{t}$ both start from $\ket{0}$, the CNOT does nothing and can be removed. 
Thus the output states are
\begin{enumerate}
    \item $\ket{0} \otimes \ket{0},\ p=(1-\delta)^{2}$;
    \item $\ket{1} \otimes \ket{1},\ p=\delta(1-\delta)$;
    \item $\ket{1} \otimes \ket{0},\ p=\delta(1-\delta)$;
    \item $\ket{0} \otimes \ket{1},\ p=\delta^{2}$.
\end{enumerate}

It is now clear that $s_{1}$ and $s_{t}$ become correlated, since the probability of them being in the same state (case 1 and 2) is higher than being in different states (case 3 and 4), for all $0 \leq \delta < 1/2$.
When $s_1$ is in $\ket{0}$, $s_t$ is more likely to be in $\ket{0}$ (case 1) than in $\ket{1}$ (case 4), so we keep this purified spin $s_{t}$ in the second step of \cref{fig_BCS}.
When $s_1$ is in $\ket{1}$, $s_t$ is equally likely to be in $\ket{0}$ or $\ket{1}$ (case 2 and 3), correspond to the completely mixed state and an effective temperature $T \rightarrow \infty$.
In this case $s_{t}$ has been heated up, so we can restore its original state by swapping $s_t$ and $s_2$ in the second step of \cref{fig_BCS}.
Overall, the density matrix of $s_{t}$ becomes closer to $\ket{0}$ at the output end.

We can calculate the exact reduced state of $s_{t}$ at the output, by taking an average over the aforementioned 4 cases (after applying the CSWAP step).
Denote the probability of being in $\ket{1}$ for $s_t$ after a BCS round as $\delta_{t}'$.
For cases 1 and 4, $s_{1}$ is in $\ket{0}$ so no SWAP gate is applied, and the probability of $s_{t}$ being in $\ket{1}$ is 0 and 1, respectively.
For cases 2 and 3, $s_{t}$ is swapped with $s_{2}$, so the probability of $s_{t}$ being in $\ket{1}$ is restored to $\delta$ in both cases.
Averaging over these 4 cases yields
\begin{equation}
    \delta_{t}' = \delta^{2} + 2 \delta^{2}(1-\delta) = 3 \delta^{2} - 2 \delta^{3} \leq \delta,\ \forall\  0 \leq \delta < \frac{1}{2}.
\end{equation}
For small $\delta$, it is reduced to order $\mc{O}(\delta^{2})$.
In the more general case where the initial $\delta$'s on each spin can be different, we have
\begin{equation}\label{eqn_post_BCS_error}
    \delta_{t}' = \delta_{1} \delta_{t} + \delta_{2} \delta_{t} + \delta_{1} \delta_{2} - 2 \delta_{1} \delta_{t} \delta_{2}.
\end{equation}
If we further assume that the ancillas $s_{1}$ and $s_{2}$ can relax much faster to their original states than $s_{t}$, so that we can effectively repeat this BCS round (where $s_{t}$ now has error $\delta_{t}'$), then $s_{t}$ can be further purified.
In the limit of performing infinitely many rounds, $s_{t}$ arrives at the steady state, whose error can be calculated by setting $\delta_{t}' = \delta_{t} = \delta_{t}^{\infty}$ in \cref{eqn_post_BCS_error}, which gives 
\begin{equation}\label{eqn_BCS_infty}
    \delta_{t}^{\infty} = \frac{ \delta_{1} \delta_{2}}{1- \delta_{1} - \delta_{2}+ 2 \delta_{1} \delta_{2}}.
\end{equation}

\subsection{MBAC}\label{sec_mbac}
Based on the previous analysis, one can see that the second CSWAP gate (and consequently, $s_2$ as well) is not needed if we can learn the state of $s_1$ after the first CNOT gate.
If $s_1$ is in $\ket{0}$, we know that $s_t$ has a higher probability of being in $\ket{0}$.
If we discard the cases where $s_1$ is in $\ket{1}$ and only keep the ones where it is in $\ket{0}$, we can reduce the error $\delta$ on the target qubit.
The working principle behind this method is similar to an error \emph{detecting} code, where the effective noise level is reduced by accepting the cases where no error occurs, and discarding those with an error occurring.

\begin{figure}[ht]
\centering
\includegraphics[width=0.6\columnwidth]{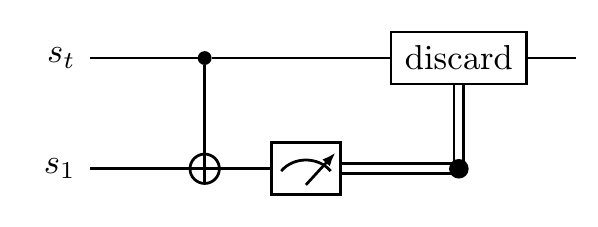}
\caption{A circuit for 2-qubit measurement-based algorithmic cooling which increases the bias of the target qubit $s_t$.
The double-line notation means ``controlled on classical outcome'', i.e., $s_{t}$ is kept for further computations when the measurement outcome is $0$ and discarded if the outcome is $1$.} 
\label{fig_MBAC}
\end{figure}

Referring to the circuit in \cref{fig_MBAC}, measuring $\ket{0}$ on $s_1$ updates the state of $s_t$ to
\begin{equation}\label{eqn_rhot}
    \rho_t' = \frac{\bra{0}_1  \tau \ket{0}_1}{\bra{0} \Tr_t [\tau] \ket{0}}
\end{equation}
where, throughout this work, we'll use $\tau$ to denote the state of the full system right before the measurement.
Starting from two states with initial error rates $\delta_1$ and $\delta_t$, the final error rate on $s_t$ upon measuring $\ket{0}$ on $s_1$ becomes
\begin{equation}\label{eqn_post_MBAC_error}
    \delta_{t}' = \frac{\delta_{1}\delta_{t}}{1-\delta_{1}-\delta_{t}+2\delta_{1}\delta_{t}}.
\end{equation}
Then, taking $\delta_t = \delta_{1} = \delta$ again for simplicity, we see that in the small-$\delta$ limit, the error is also reduced to $\mc{O}(\delta^{2})$ to leading order. 
Importantly, the target qubit $s_{t}$ is not being measured, which resolves the first issue we raised in \cref{sec_intro}.
Assuming that the ancillary qubit $s_1$ can maintain sufficiently long coherence, we can store them in the quantum computer and measure them at the end of the computation, along with other qubits.
We then post-select only those measurement outcomes where $s_{1}$ measures to $0$.
Interestingly, comparing with \cref{eqn_BCS_infty}, we see that applying the above protocol once achieves the same polarization on $s_{t}$ as applying infinitely many rounds of BCS, if initially $\delta_{t} = \delta_{2}$.

The above 2-qubit protocol, which we'll call MBAC-2, forms a basis for analyzing expanded versions of MBAC.
For conventional AC protocols there are two major ways to expand them, either by using more ancillary qubits, or by repeating the protocol for multiple rounds.
Since we have assumed that the ancillary qubits cannot be reused once they have been measured, MBAC protocols cannot be expanded by repeating it for multiple rounds.
However, it is feasible to use more ancillary qubits to achieve better cooling.
Specifically, imagine expanding $s_{1}$ to $k-1$ qubits in \cref{fig_MBAC}.
We apply $k-1$ CNOT gates controlled by $s_{t}$ and targeted on $s_{i},\ i=1,...,k-1$, and measure all $s_{1},...,s_{k-1}$ at the end.
The target $s_{t}$ is kept only if \emph{all} measurement outcomes are $0$, and is discarded otherwise.
We call the above protocol MBAC-$k$, which serves as an expanded version of MBAC-2.

To analyze MBAC-$k$, observe that it is equivalent to repeating $k-1$ ``rounds'' of MBAC-2, if the ancilla in MBAC-2 is allowed to return to its original state after being measured and can be reused again. 
The evolution of noise in a single ``round'' is already given in \cref{eqn_post_MBAC_error}, so the evolution for MBAC-$k$ can be recursively calculated from \cref{eqn_post_MBAC_error}.
Furthermore, by assuming again $\delta_t = \delta_{1} = \delta$ initially and observing the first few terms, we can infer the analytic solution for this series, given by
\begin{equation}\label{eqn_delta_k}
    \delta_{t}[k] = \frac{\delta^{k}}{\delta^{k}+(1-\delta)^{k}},
\end{equation}
where from now on we use $\delta_{t}[k]$ to denote ``SP error on $s_{t}$ after applying MBAC-$k$''.
In particular, $\delta_{t}[1]$ will be used to denote the initial SP error on $s_{t}$.
The solution in \cref{eqn_delta_k} can be readily verified by plugging it back into \cref{eqn_post_MBAC_error}.
From this solution one can see how $\delta_{t}$ decreases approximately exponentially in $k$, especially in the small-$\delta$ limit.

\begin{figure}[ht]
\centering
\includegraphics[width=1.0\columnwidth]{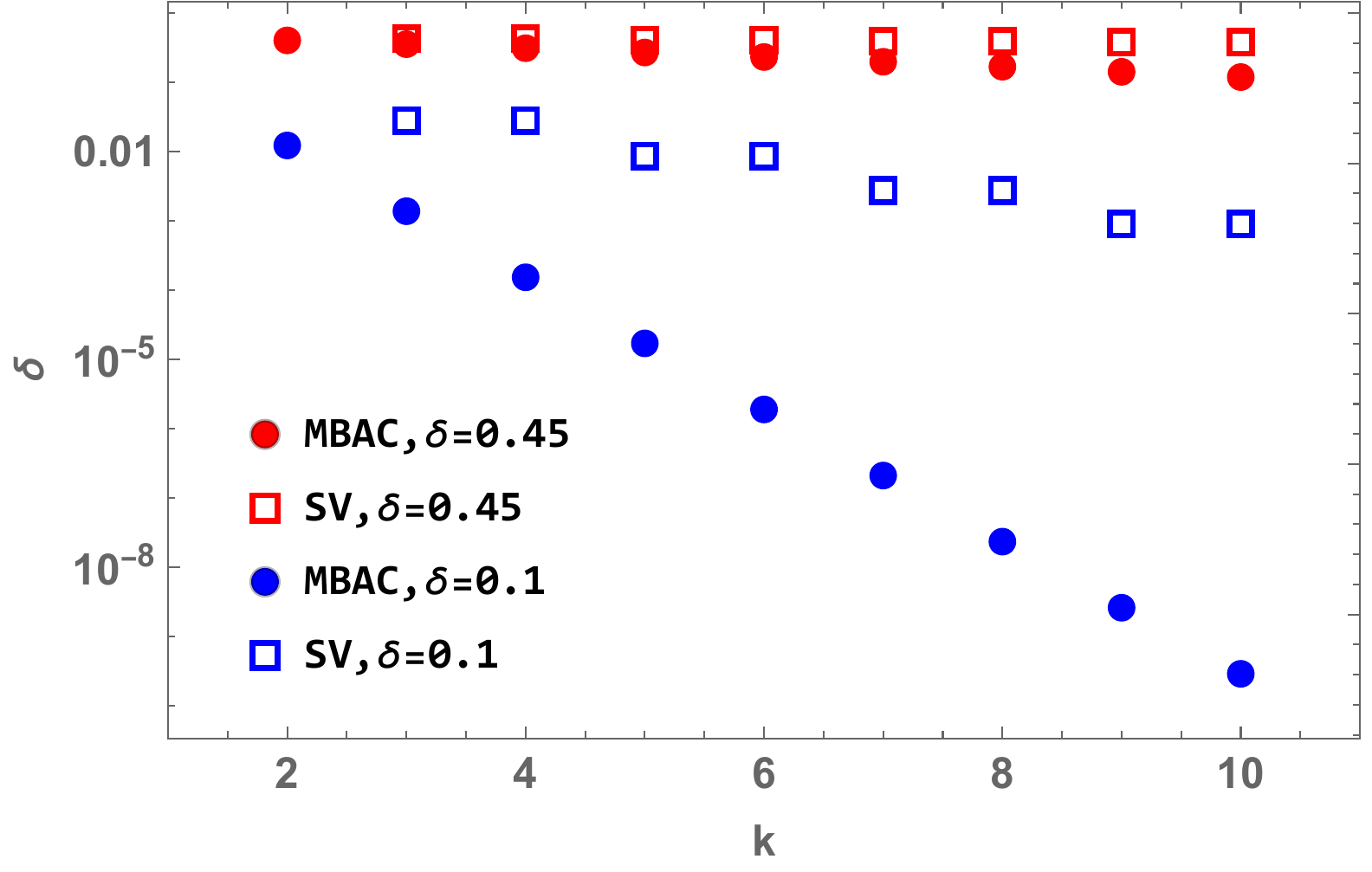}
\caption{Simulated evolution of $\delta_{t}$ between MBAC-$k$ and SV-$k$, plotted on semi-log scale. The initial error (red: $\delta=0.45$; blue: $\delta=0.1$) are assumed to be the same on all qubits. The effect of decoherence is assumed negligible and gates are assumed to be ideal.}
\label{fig_comparison}
\end{figure}

To make a fair comparison between MBAC and conventional AC, we now briefly review the reversible scheme first proposed by Schulman and Vazirani~\cite{Schulman1999}.
The main step in reversible AC is called entropy compression, where through a unitary map $\mc{U}$, the entropy is extracted from some subset of qubits and transferred to another subset of qubits.
If we constrain the system to start and end in diagonal states, and the goal is to cool down only one qubit, then it can be shown that the optimal unitary $\mc{U}$ is to perform a descending sort on the diagonal elements of the full system's density matrix~\cite{Rodriguez-Briones2016,Schulman2005Physical}.
We thus consider this to be an upper bound on the performance of Schulman-Vazirani type of cooling, and will call this scheme SV-$k$ if it uses a total of $k$ qubits to cool one qubit.
In \cref{fig_comparison}, we compared MBAC-$k$ (in circles) and SV-$k$ (in squares) starting from two initial noise levels, $\delta=0.1$ (in blue) and $\delta = 0.45$ (in red).
The advantage of allowing projective measurement into the task of cooling, compared to the optimal reversible scheme, is clearly visible.

\subsection{SPAM error characterization}\label{sec_SPAM_char}
In \cref{sec_review_AC}, we have already seen how the parameter $\delta$ can be interpreted as a measure of state preparation (SP) error.
We will now formally define what SP and M errors are, then demonstrate how MBAC can be used to characterize SPAM errors.
Our definition here will closely follow the one given in~\cite{Lin2021}.
The ideal SPAM operators are denoted by a density operator $\rho$ and a $2$-outcome POVM $M = \{M_0,M_1\}$, which satisfy the physicality constraints that $\rho$ and $M_i$ are positive-semidefinite Hermitian operators, and $M_0 + M_1 = \sigma_{I}$.
We assume that ideally, $\rho = M_0 =\ketbra{0}{0}$.
The SPAM error is defined as the probability of obtaining an incorrect outcome when measuring the initial state: that is,
\begin{equation}\label{eqn_del_SPAM_defn}
    \delta_{\text{SPAM}} \coloneqq 1- \Tr[\rho M_0].
\end{equation}
We now define the SP error, denoted as $\delta_{\text{SP}}$, to be equal to $\delta_{\text{SPAM}}$ with an ideal measurement operator.
Similarly, the M error (denoted as $\delta_{\text{M}}$) is defined as $\delta_{\text{SPAM}}$ with an ideal input state.
While the total SPAM error $\delta_{\text{SPAM}}$ is a measurable quantity, $\delta_{\text{SP}}$ and $\delta_{\text{M}}$ are not, when both state preparation and measurement errors are present.
However, one may design algorithms that allow one to isolate the contributions to the total error from state preparation or measurement processes.

We now demonstrate how one can separately estimate $\delta_{\text{SP}}$ and $\delta_{\text{M}}$, based on the concept of AC.
For simplicity, we first assume that both the imperfect state and measurement operators are of the same form
\begin{equation}\label{eqn_SPAM_simplified}
    \rho = \begin{pmatrix} 1-\delta_{\text{SP}} & 0 \\ 0 & \delta_{\text{SP}} \end{pmatrix},\ M_0 = \begin{pmatrix} 1-\delta_{\text{M}} & 0 \\ 0 & \delta_{\text{M}} \end{pmatrix},
\end{equation}
so that the total SPAM error is given (from \cref{eqn_del_SPAM_defn}) by
\begin{equation}\label{eqn_SPAM_SP_M}
    \delta_{\text{SPAM}} = \delta_{\text{SP}} + \delta_{\text{M}} - 2 \delta_{\text{SP}} \delta_{\text{M}}.
\end{equation}

As before, we will also assume that $\delta_{\text{SP}}, \delta_{\text{M}} \in [0,0.5)$.
Our goal is to separately estimate $\delta_{\text{SP},t}$ and $\delta_{\text{M},t}$ on $s_{t}$.
For now we also assume, for simplicity, that measurement operations on all \emph{ancillary} qubits are \emph{ideal}, and the only noisy measurement is the one on $s_{t}$.
This allows us to directly apply the previous calculations.
This assumption will be relaxed in the next section.

Since $\delta_{\text{SPAM},t}$ can be obtained from directly measuring $s_{t}$, the problem of separately characterizing SPAM is then reduced to estimating either $\delta_{\text{SP},t}$ or $\delta_{\text{M},t}$: from the symmetry between the two, once either is known, the other can also be calculated from the second equation.

It is now possible to intuitively see how AC can be used to resolve SPAM errors.
We saw in \cref{eqn_delta_k} and \cref{fig_comparison} that MBAC can quickly reduce the error $\delta_{t}$ on $s_{t}$ close to $0$.
Now imagine two separate experiments where in the first one, we measure $s_{t}$ directly and obtain $\delta_{\text{SPAM},t}$.
In the second one, we first apply multiple rounds of MBAC until the final bias on the target is sufficiently close to $1$ (we'll later show how this can be determined), then measure the target qubit.
Since the measurement operation is independent of the qubit state, $\delta_{\text{M},t}$ is directly obtained from the measurement result, since the input state is now ideal.
From here, $\delta_{\text{SP},t}$ can be easily calculated from \cref{eqn_SPAM_SP_M}.
We have thus separately estimated SPAM errors by first eliminating the SP error, determining the measurement error, then inferring the SP error from the total $\delta_{\text{SPAM},t}$.

We have so far focused on the case of diagonal state and measurement operators.
To justify this, below we describe an averaging technique to convert arbitrary 1-qubit SPAM elements to this simpler case.
Begin by noting that we can generally write
\begin{equation}\label{eqn-parametrization}
    \begin{gathered}
    \rho = \frac{1}{2} (\sigma_{I}+s_{x} \sigma_{x} + s_{y} \sigma_{y} + s_{z} \sigma_{z})\\
    M_0 = \frac{1}{2} (m_{i} \sigma_{I} + m_{x} \sigma_{x} + m_{y} \sigma_{y} + m_{z} \sigma_{z})
    \end{gathered}
\end{equation}
where the $s$'s and $m$'s are unknown parameters ($s_{i} = 1$ because $\Tr[\rho] = 1$).
Assuming ideal quantum gates, we can obtain an effective initial state with $s_{x} = s_{y} = 0$ for an arbitrary 1-qubit circuit as follows.
We perform two separate experiments, where in the first we apply the original circuit, and in the second we apply a $\sigma_{z}$ gate immediately after the state preparation, then carry out the same circuit.
Due to linearity of quantum operations, the average of measurement outcomes from the two experiments is then equivalent to one where $\rho_{\text{eff}} = \frac{1}{2} (\rho + \sigma_{z} \rho \sigma_{z}^{\dagger}) = \frac{1}{2} (\sigma_{I} + s_{z} \sigma_{z})$.
Similarly, we can also make $m_{x} = m_{y} = 0$ by averaging the results from the original experiment and one where a $\sigma_{z}$ gate is applied immediately before the measurement.
To set $m_{i} = 1$, we can average the original circuit with one where a $\sigma_{x}$ is applied immediately before the measurement, and the outcomes $0$ and $1$ are relabelled (so that outcome $0$ corresponds to the POVM element $M_1$ and \textit{vice versa}).
This reduces the SPAM operators to the ones described by \cref{eqn_SPAM_simplified}.

\subsection{MBAC with measurement errors}\label{sec_measurement_error}
We now deal with the second issue raised in \cref{sec_intro}, and study the performance of MBAC when the measurement error is non-zero on all \emph{ancillary} qubits as well, thereby relaxing the assumption made in the previous section.
We will show that MBAC still performs well if $\delta_{\text{SPAM}}$ on each ancilla is not very large, thereby relaxing the most crucial assumption in SPAM characterization.
Again, in this section we will assume that quantum gates are ideal, which means that the SPAM averaging processes are also ideal, and the state and measurement operators can each be described by a single parameter.

The problem setup is as follows.
We assume that each qubit $i$ in a quantum computer has an independent state preparation error $\delta_{\text{SP},i}$, and measurement error $\delta_{\text{M},i}$, where $i$ labels the qubit.
The target qubit is labelled by $t$ as usual.
The goal is again to learn $\delta_{\text{SP},t}$ and $\delta_{\text{M},t}$.
But as we show later, similar arguments can be used to learn all $\delta_{\text{SP},i}$ and $\delta_{\text{M},i}$ if desired.

Consider performing one successful round of the MBAC-2 protocol with two qubits $s_{t}$ and $s_{1}$.
With an imperfect measurement, we generalize the case of projective measurement in \cref{eqn_rhot} to a POVM measurement, so that performing one successful round of MBAC updates the state of $s_t$ to~\cite{watrous2018theory}
\begin{equation}\label{eqn_POVM_rho_prime}
    \rho_{t}' = \frac{\Tr_{1} [\tau (I \otimes M_0)]}{\Tr[\Tr_{t}[\tau] M_0]},
\end{equation}
where, again, $\tau$ denotes the state of the full system immediately before measurement.
Using \cref{eqn_POVM_rho_prime}, following again the circuit in \cref{fig_MBAC} and the parametrization in \cref{eqn_SPAM_simplified}, we calculate the SP-error on $s_t$ after one round of MBAC to be
\begin{equation}\label{eqn_del_SP_general}
\begin{aligned}
    \delta_{\text{SP},t}' &=\delta_{\text{SP},t} \frac{2(\delta_{\text{SP},1} +\delta_{\text{M},1}-2\delta_{\text{SP},1} \delta_{\text{M},1})}{1+(1-2\delta_{\text{SP},1})(1-2\delta_{\text{SP},t})(1-2\delta_{\text{M},1})}\\
    &= \delta_{\text{SP},t} \frac{2\delta_{\text{SPAM},1} }{1+(1-2\delta_{\text{SP},1})(1-2\delta_{\text{SP},t})(1-2\delta_{\text{M},1})}
\end{aligned}
\end{equation}
where the second equality comes from \cref{eqn_SPAM_SP_M}.
The ratio
\begin{equation}\label{eqn_defn_r}
    \frac{\delta_{\text{SP},t}}{\delta_{\text{SP},t}'} = \frac{1+(1-2\delta_{\text{SP},1})(1-2\delta_{\text{SP},t})(1-2\delta_{\text{M},1})}{2\delta_{\text{SPAM},1} }
\end{equation} is a measure of the improvement of SP-error on $s_t$ after one round of MBAC, which is better when larger.
For example, a ratio of $100$ implies that $\delta_{\text{SP},t}$ has been reduced by a factor of $100$.
Intuitively, the improvement should be more significant when there is less error on $s_1$: indeed, if $\delta_{\text{SP},1} = \delta_{\text{M},1} = 0$ in \cref{eqn_del_SP_general}, then $\delta_{\text{SP},t}' = 0$ and $s_t$ will always be projected to $\ketbra{0}{0}$ when the measurement outputs $0$ on $s_1$.
In the more general case where SPAM error on $s_1$ is present, we observe that the numerator on the RHS of \cref{eqn_defn_r} is always $\geq 1$, in the relevant region where $\delta_{\text{SP},1}, \delta_{\text{SP},t}, \delta_{\text{M},1} \in [0,1/2)$.
Therefore,
\begin{equation}\label{eqn_r_bound}
    \frac{\delta_{\text{SP},t}}{\delta_{\text{SP},t}'} \geq \frac{1}{2 \delta_{\text{SPAM},1}}.
\end{equation}
Furthermore, in the limit where all the $\delta_{\text{SP},1}$, $\delta_{\text{SP},t}$, $\delta_{\text{M},1} \ll 1$, the bound in \cref{eqn_r_bound} can be improved by approximately a factor of $2$ to simply $1/\delta_{\text{SPAM},1}$.

Next, recall from \cref{sec_mbac} that MBAC-$k$ is equivalent to repeating $k-1$ rounds of MBAC-2.
Using mathematical induction, we see that by applying one successful run of MBAC-$k$, the final SP-error on $s_t$ is upper bounded by
\begin{equation}\label{eqn_delta_bound}
    \delta_{\text{SP},t}[k] \leq \delta_{\text{SP},t} \prod_{i=1}^{k-1}  (2\delta_{\text{SPAM},i}).
\end{equation}
Recall that $\delta_{\text{SPAM},i}$ is a measurable quantity obtained by measuring the initial state on ancilla $s_{i}$.
The product simply corresponds to the probability of directly measuring all ancillary qubits $s_1 \dots s_{k-1}$ after they are prepared, and getting the output 1 on all qubits.
Therefore, given $\delta_{\text{SPAM},i}$ on each qubit (which can be obtained before the experiment, during the calibration step), \cref{eqn_delta_bound} guarantees that $\delta_{\text{SP},t}$ from the output of a successful run of MBAC-$k$ is at least reduced by $\prod_{i=1}^{k-1}  (2\delta_{\text{SPAM},i})$.
This shows that finite measurement error on the ancillary qubits do not pose a fundamental limitation to the cooling power of MBAC.
In particular, as long as each $\delta_{\text{SPAM},i} < 1/2$ (which we will assume from now on), the output state is guaranteed to be more pure than the input.
Moreover, if both $\delta_{\text{SP},t}$ and all $\delta_{\text{SPAM},i}$'s are upper bounded by a constant $\delta$, then $\delta_{\text{SP},t}[k]$ is simply upper bounded by $\delta^{k}$.

\subsection{Number of trials needed in MBAC-$k$}\label{sec_trials_needed}
Next, we study whether the probabilistic nature of MBAC hinders its usefulness in practice.
We will see that it remains practically useful for a wide range of experimentally relevant SPAM error rates.
To illustrate the problem, first ignore any measurement error and consider a run of MBAC-2.
The post-measurement state upon measuring $1$ (which imply a failed run) can be computed by changing all $0$'s to $1$'s in \cref{eqn_rhot}, leading to
\begin{equation}\label{eqn_failure_eps}
    \delta_{t,\text{fail}}' = \frac{\delta_{t} (1- \delta_{1})}{\delta_{t} + \delta_{1} - 2 \delta_{t}\delta_{1}} = \frac{1}{2} - \frac{\delta_{1}-\delta_{t}}{2(\delta_{t} + \delta_{1} - 2 \delta_{t}\delta_{1})}.
\end{equation}
One sees that, if $\delta_{1} = \delta_{t}$ initially, then obtaining measurement outcome $1$ on $s_{1}$ will heat up the state to a completely mixed one.
The probability of failure is
\begin{equation}
    p_{\text{fail}} = \delta_{1} + \delta_{t} - 2 \delta_{1}\delta_{t}.
\end{equation}
For MBAC-$k$, since we need all the $k-1$ measurements to succeed, the rate of success decreases exponentially with $k$, and it becomes increasingly difficult to get a successful run.

Fortunately, recall from \cref{eqn_delta_bound} that the target polarization also improves exponentially fast with increasing number of ancillas.
Therefore, the intuition is that one does not need a large $k$ value to achieve significant cooling, which in turn will not have a vanishingly small success probability.
Below we'll make this intuition more concrete.
Specifically, we ask the following question: if we would like to reduce $\delta_{\text{SP},t}$ by a factor of $r$, i.e., we want $\delta_{\text{SP},t}/\delta_{\text{SP},t}[k] = r$, how many runs are needed to achieve this cooling ratio?
To answer this we will start from \cref{eqn_delta_bound}, and first derive a relation between $k$ and $r$.
Note that because each $\delta_{\text{SPAM},i}$ can be different, the most general expression will involve all $\delta_{\text{SPAM},i}$'s.
In order to obtain an expression in $k$, we now make the assumption that all $k-1$ ancillary qubits have the same $\delta_{\text{SPAM}}$, which we will denote as $\delta_{\text{SPAM},a}$ where $a$ stands for the word ``ancillary''.
The case where the $\delta_{\text{SPAM},i}$'s are different can be bounded similarly by setting $\delta_{\text{SPAM},a}$ to the highest among all $\delta_{\text{SPAM},i}$.

With this assumption, we obtain the following inequality between $k$ and $r$:
\begin{equation}\label{eqn_r_and_k}
    r = \frac{\delta_{\text{SP},t}}{\delta_{\text{SP},t}[k]} \geq (2 \delta_{\text{SPAM},a})^{-(k-1)}.
\end{equation}
Taking the logarithm of both sides and rearranging the terms (note that $\log(2\delta_{\text{SPAM},a}) < 0$) results in
\begin{equation}\label{eqn_k_bound}
    k-1 \leq \frac{\log(r)}{-\log(2\delta_{\text{SPAM},a})}.
\end{equation}
This allows the experimentalist to determine the total number of ancillary qubits needed in order to cool to the desired noise level, based on their hardware specifications (i.e., the $\delta_{\text{SPAM},a}$ on their hardware).
Importantly, this upper bound scales logarithmically with $r$.

Next, we compute the expectation value of the total number of runs needed before having a successful run, in order to achieve a cooling ratio $r$.
We show in \cref{sec_appen_delta_bound} that the expected number of runs is upper bounded by a function, 
\begin{equation}\label{eqn_N_upper_bound}
    N_{\text{upper}}(r) = (r)^{\frac{\log(A)}{\log(B)}}
\end{equation}
where
\begin{equation}
    \begin{gathered}
        A = (1-\delta_{\text{SP},t}[1] - \delta_{\text{SP,a}} + 2\delta_{\text{SP},t}[1] \delta_{\text{SP,a}})(1- \delta_{\text{M},a})\\
        B = 2 \delta_{\text{SPAM},a}.
    \end{gathered}
\end{equation}
We see that \cref{eqn_N_upper_bound} scales polynomially in $r$.
Note that $A$ is simply the success probability (i.e., measuring $0$) of doing the first run of MBAC-2, so the exponent is a measurable quantity.
Thus, the upper bound can be calculated given a target $r$.

To understand the behavior of $N_{\text{upper}}(r)$ more concretely, we can further simplify \cref{eqn_N_upper_bound} by assuming again $\delta_{\text{SP},t}[1] =  \delta_{\text{SP,a}} \coloneqq \delta_{\text{SP}}$, i.e., both the target and the ancillas have the same initial error $\delta_{\text{SP}}$.
The exponent then becomes (recall $\delta_{\text{SPAM}}$ from \cref{eqn_SPAM_SP_M})
\begin{equation}\label{eqn_logA_logB}
    \frac{\log(A)}{\log(B)} = \frac{\log((1-2\delta_{\text{SP}} + 2 \delta_{\text{SP}}^2)(1-\delta_{\text{M},a}))}{\log(2(\delta_{\text{SP}}+\delta_{\text{M},a}-2\delta_{\text{SP}}\delta_{\text{M},a}))}.
\end{equation}

\begin{figure}[ht]
\centering
\includegraphics[width=1.0\columnwidth]{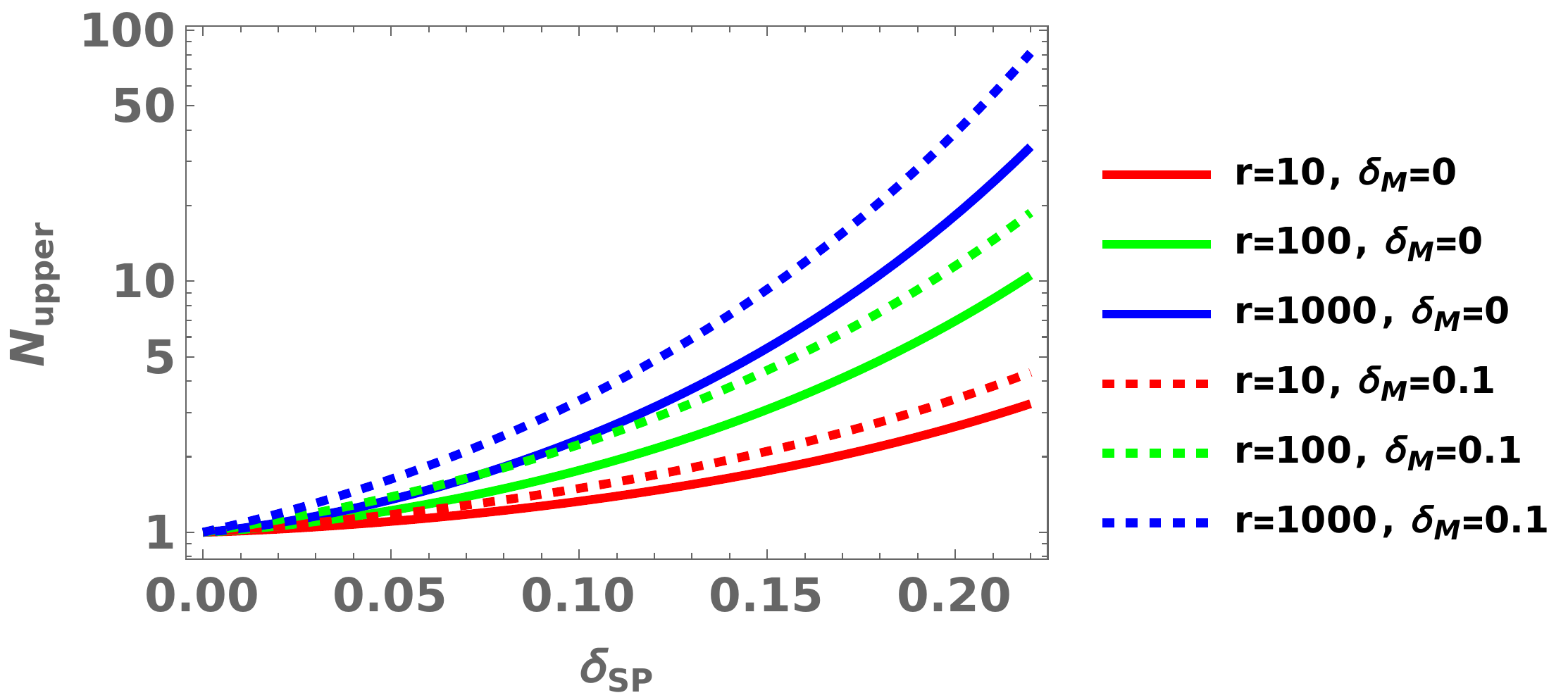}
\caption{Upper bound $N_{\text{upper}}$ on expected number of runs required for different values of $r$ (the improvement ratio defined in \cref{eqn_r_and_k}), versus the initial SP error $\delta_{\text{SP}}$ (assumed to be the same on the target and all ancillary qubits). Solid and dashed lines represent the cases of $\delta_{\text{M},a}=0$ and $\delta_{\text{M},a}=0.1$, respectively.}
\label{plot_appen_scaling}
\end{figure}

Shown in \cref{plot_appen_scaling} are plots of $N_{\text{upper}}(r)$ in \cref{eqn_N_upper_bound} as a function of $\delta_{\text{SP}}$, after making the simplifying assumption in \cref{eqn_logA_logB}, for a few chosen values of $r$ and $\delta_{\text{M},a}$.
As the initial SP error rate approaches $0.5$ (the theoretical maximum), MBAC fails since the measurement is simply giving random outputs.
In this case, $N_{\text{upper}}(r)$ diverges as expected.
On the other hand, for reasonably low values of $\delta_{\text{SP}}$, we see that the expected number of runs before achieving a successful one is rather low.
For example, for $\delta_{\text{SP}} = 0.1$, $\delta_{\text{M},a}=0$, and $r=1000$, we expect to obtain a successful run in about $2$ trials.
For $\delta_{\text{SP}} = 0.1$, $\delta_{\text{M},a}=0.1$, and $r=1000$, the expected number of trials is approximately $3$.
Note that this corresponds to a case of $10\%$ SP error rate plus $10\%$ measurement error rate on the ancillas, combining to almost $20\%$ of total SPAM error rate.
Many modern quantum computing platforms~\cite{Shadbolt2012,walter2017rapid,Wright2019} can now achieve SPAM error rates below this level.
In these cases, MBAC will be a useful and easy tool to significantly improve the quality of state preparation in quantum computers.

\section{Discussion}\label{sec_conclusion}
We introduce a new variant of algorithmic cooling (AC) protocol based on the ability to perform imperfect measurements on individual qubits, which we call measurement-based AC (MBAC).
Using this method, we develop a novel and simple way to separately characterize state preparation and measurement errors, by eliminating the former using MBAC and directly obtaining the latter.
Our approach is applicable to many current quantum computing platforms, and significantly outperforms the optimal reversible AC protocol in the absence of measurement errors.
Moreover, its cooling ability can still be lower bounded by some experimentally measurable quantities when measurement errors are present.
Despite the probabilistic nature of MBAC, we have shown that it remains practically useful for many current quantum processors.
We believe our method can be a helpful tool for benchmarking and improving current NISQ-era quantum computers, and can provide further insights to measurement as a resource for cooling.

A few open questions remain after our work.
First, while we have shown that measurement errors do not pose a fundamental challenge for MBAC, it remains an interesting question to also study the effect of errors in the quantum gates used \cite{Ben-Or2013,Lin2021}.
Second, it would be useful to investigate the performance of MBAC with fully general SPAM operators, without the averaging techniques used in our work.
Third, it can be very interesting to integrate the concept of MBAC into other algorithmic cooling protocols (e.g.,~\cite{boykin2002algorithmic,Fernandez2004Algorithmic,Schulman2005Physical,Elias2011,Rodriguez-Briones2017,Rodriguez-Briones2017a,Alhambra2019}; also see the reviews \cite{Brassard2014,Park2016}), which goes far beyond the BCS subroutine mentioned in~\cref{sec_mbac}, to develop new protocols that may have better performances than these current protocols.
Last, our work suggests that measurements may be regarded as a resource in quantum thermodynamic theories.
It would be interesting to explore the possibility of integrating measurement operations into the current quantum resource theory framework.

    


\section{Methods}
\subsection{Calculating expected number of runs before a successful one}\label{sec_appen_delta_bound}
Here we calculate the expected number of failed MBAC-$k$ runs before having a successful one.
The scenario is that one will continue running the experiment, until a successful round of MBAC-$k$ occurs.
Assume that one would like to reduce $\delta_{\text{SP},t}$ by a factor of $r$ (e.g., $r=100$ or $r=1000$ can be set by the experimentalist), by using MBAC-$k$.
According to Eq. (14) in the main text, the final SP error is exponentially suppressed in terms of successful cooling rounds.

We again think about MBAC-$k$ as an $(k-1)$-step protocol, where each step corresponds to measuring an ancilla and tracing it out after the measurement.
The probability of measuring $0$ on the $i$-th step is
\begin{equation}\label{eqn_appen_p_0i}
    p_{0,i} = (1-\delta_{\text{SP},t}[i] - \delta_{\text{SP,a}} + 2\delta_{\text{SP},t}[i] \delta_{\text{SP,a}})(1- \delta_{\text{M},a}).
\end{equation}
The measurements are independent for each round, so the probability of measuring all $0$'s is given by the product of $p_{0,i}$ from each round.
Since $p_{0,i}$ improves as $i$ increases, we can lower bound all of them by $p_{0,1}$, i.e., the probability of getting $0$ in the first run of MBAC-2.

Thus, we see that the expected total number of runs is upper bounded by a simpler case, where $n$ independent Bernoulli trials are conducted in series, each having success probability $p_{0,1}$.
The probability of having all trials successful, which we will call the success probability, is $(p_{0,1})^{k-1}$.
Note that for a Bernoulli trial that has success probability $p$, the expected number of tests $n$ can be calculated to be
\begin{equation}
    \begin{aligned}
    \mathds{E}(n) =& \sum_{i} n_{i} \cdot p_{i}(n_{i})\\
    = & 1\cdot p + 2p(1-p) + 3p(1-p)^2+ \dots\\
    = & p\sum_{n=0}^{\infty} (n+1)(1-p)^n \\
    = & \frac{p}{p^2} = \frac{1}{p}
    \end{aligned}
\end{equation}
as expected.
Thus, if we denote the expected number of tests \emph{in reality} by $N$, we have
\begin{equation}\label{eqn_appen_N_bound}
    N \leq \frac{1}{(p_{0,1})^{k-1}},
\end{equation}
where $p_{0,1}$ is defined in \cref{eqn_appen_p_0i}.
Next, recall that \cref{eqn_k_bound} gives an upper bound on the required number of ancillary qubits, $k-1$.
Because $p_{0,1} \in (0,1)$, $p_{0,1}^{x}$ decreases with $x$ for $x \geq 0$, so that $(1/p_{0,1})^{x}$ increases with $x$ for $x \geq 0$.
Combining this with \cref{eqn_appen_N_bound}, we have
\begin{equation}\label{eqn_appen_A_B}
    N \leq \left(\frac{1}{p_{0,1}}\right)^{k-1} \leq \left(\frac{1}{p_{0,1}}\right)^{\frac{\log(r)}{-\log(2\delta_{\text{SPAM},a})}} \coloneqq A^{\frac{\log(r)}{\log(B)}}
\end{equation}
where
\begin{equation}
    A = p_{0,1},\ B = 2 \delta_{\text{SPAM},a}.
\end{equation}
We now define the expression on the RHS of \cref{eqn_appen_A_B} as an upper bound on $N$, i.e.,
\begin{equation}\label{eqn_appen_N_upper}
    N_{\text{upper}}(r) \coloneqq A^{\frac{\log(r)}{\log(B)}}.
\end{equation}
Note that this upper bound is only a function of $r$ and some measurable quantities that are specific to the machine.
Taking the log on both sides of \cref{eqn_appen_N_upper} gives
\begin{equation}
    \log(N_{\text{upper}}(r)) = \frac{\log(r)}{\log(B)}\log(A) = \log(r) \frac{\log(A)}{\log(B)}
\end{equation}
and taking the exponential on both sides again gives
\begin{equation}
    N_{\text{upper}}(r) = (r)^{\frac{\log(A)}{\log(B)}}
\end{equation}
which is polynomial in $r$, as claimed.

\section{Data availability}
The data supporting the findings of this work are available from the corresponding author upon request.

\bibliography{SPAM-AC}

\section{Acknowledgments}
R.L. acknowledges funding from Mike and Ophelia Lazaridis.
R.L., J.L., and T.M. thank the Schwartz/Reisman Foundation. 
J.L. is supported by NSERC Canada. 
T.M. was also partially supported by Israeli MOD.

\section{Author Contributions}
All authors contributed extensively to the presented work. 
R.L. and T.M. conceived the original ideas and supervised the project.
J.L. performed analytical studies and numerical simulations, and generated different versions of the manuscript.
R.L. and T.M. verified the calculations, proof-read the manuscript, and applied many insightful updates.

\end{document}